\documentclass[tightenlines,secnumarabic,amssymb,amsmath,nobibnotes,aps,prd,nofootinbib]{revtex4}

\usepackage{amsbsy}
\usepackage{amssymb}
\usepackage{amsmath}
\usepackage{graphicx}

\newcommand{\be}{\begin{equation}}
\newcommand{\ee}{\end{equation}}
\newcommand{\bear}{\begin{eqnarray}}
\newcommand{\eear}{\end{eqnarray}}

\newcommand{\z}{\hat{z}}

\newcommand{\cM}{{\cal M}}

\newcommand{\cR}{{\cal R}}
\newcommand{\cS}{{\cal S}}

\begin{document}

\title{The separable analogue of Kerr in Newtonian gravity}

\author{Kostas Glampedakis$^{1,2}$ and Theocharis A. Apostolatos$^{3}$ }

\affiliation{
$^{1}$Departamento de F\'{i}sica, Universidad de Murcia, E-30100 Murcia, Spain
\\
$^{2}$Theoretical Astrophysics, University of T\"ubingen, Auf der Morgenstelle 10, T\"ubingen, D-72076, Germany
\\
$^{3}$Section of Astrophysics, Astronomy, and Mechanics, Department of Physics, University of Athens, Panepistimiopolis
Zografos GR15783, Athens, Greece}
 
\date{\today}

\begin{abstract}
General Relativity's Kerr metric is famous for its many symmetries which are responsible for the separability of the
Hamilton-Jacobi equation governing the geodesic motion and of the Teukolsky equation for wave dynamics. We show that there
is a unique stationary and axisymmetric Newtonian gravitational potential that has exactly the same dual property of
separable point-particle and wave motion equations. This `Kerr metric analogue' of Newtonian gravity 
is none other than Euler's 18th century problem of two-fixed gravitating centers.
\end{abstract} 
  
\maketitle


\section{Introduction}
\label{sec:intro}

The Kerr metric describing rotating black holes, apart from its extreme physical significance is also abundantly enriched with symmetries \cite{chandrabook}. The existence of these symmetries is mirrored in the equations governing the motion of particles and waves in the Kerr gravitational field. The Hamilton-Jacobi equation for particle geodesic motion and the Teukolsky equation for wave dynamics are known to be separable, thus offering a deep understanding of the underlying physics \cite{teuk73}. Indeed, the impressive theoretical advance in black hole physics that took place in the black-hole `golden age' 1970s-1990s was based on this special property of the Kerr metric \cite{chandrabook}.   

On the other hand, Newtonian gravity of course cannot create black holes. However, some Newtonian gravitational potentials may possess many of the symmetries of the Kerr metric. For instance, the notion of a `third integral' of motion (besides the obvious integrals of
energy and angular momentum along the symmetry axis) is a well-known concept in Newtonian gravity, e.g., in the context of galactic potentials \cite{galactic}, and is associated with integrability, that is, the particle motion having as many conserved quantities in 
involution with each other as the number of degrees of freedom \cite{Arnold}. In black hole physics this integral of motion is 
known as the Carter constant~\cite{carter} and its existence is linked with the `hidden' symmetry of a rank-two Killing tensor 
in the Kerr metric~\cite{stewart} (in contrast with the more obvious symmetries of stationarity and axisymmetry
which lead to a conserved orbital energy and angular momentum along the spin axis, respectively). 

A famous example of a Newtonian potential endowed with a Carter-like constant is the Euler solution for the two-centre problem,
i.e. the gravitational field sourced by a pair of spatially-fixed point masses. A recent account of the integrability of particle motion in the Euler field and a simple derivation of its Carter constant has been provided by Lynden-Bell \cite{lyndenbell}.
More recently, Will \cite{will10} revealed an unexpected aspect of Euler's 250-year old solution.  
Within the whole class of solutions of the Poisson equation, there is a unique axisymmetric and reflection-symmetric solution 
(besides the spherically-symmetric monopole solution) which has a third integral like the Carter constant. 
This special potential is none other than the Euler solution (with equal-mass centres). Remarkably, the mass multipole moments
in which the Euler potential can be decomposed are interrelated in exactly the same way as the mass multipoles of the Kerr metric!

Based on these results one is compelled to argue that the Euler potential is the Kerr analogue in Newtonian gravity (this 
analogy was first noticed long time ago, see Refs.~\cite{keres67},\cite{israel70}). 
However, the close relationship between the two gravitational fields has been established within the context of point-particle motion, i.e. the presence of a Carter-like constant and its relation to the separability of the Hamilton-Jacobi equation~\cite{LLbook}. 
Knowing that the Kerr metric is also special with respect to wave dynamics~\cite{teuk73}, the emerging question is whether
we can push the Euler/Kerr similarity even further, by comparing these two gravitational fields with respect to wave-dynamics properties.

More specifically, a key question to be addresed is this: does the Euler problem lead to a separable wave equation in the 
context of Newtonian gravity, in the same way that the Teukolsky equation for the Kerr metric has a separable solution? 
Moreover, is this the only Newtonian potential (besides the spherically symmetric one) with the dual property of
rendering separable both the Hamilton-Jacobi {\em and} the wave equations? If the answer is affirmative, then the analogy
between the relativistic Kerr metric and the Newtonian Euler potential would be `complete'. Besides having significant theoretical value in itself, such an analogy between the two fields could be used to get a deeper insight into these two fields or even, perhaps, predict new properties for them.

In this paper we provide an affirmative answer to the above questions. That is, we show that the Euler potential is the
unique axisymmetric and reflection-symmetric Newtonian potential that allows the separability of the scalar wave 
equation (in addition to that of the Hamilton-Jacobi equation), using a spheroidal coordinate system.

The rest of the paper is organized as following: In Section~\ref{sec:euler} we discuss the occurrence of the Carter constant in 
Newtonian gravity and  the connection between the Euler potential and the Kerr metric. In Section~\ref{sec:wave} we formulate the wave equation in Newtonian gravity and discuss its separability. In this section we show, with a modicum of effort, that in usual spherical coordinates the separability of the wave equation is possible only for a monopole (i.e. spherically symmetric) potential. 
Within the more general class of prolate/oblate spheroidal coordinates though, the wave equation is separable if and only if the potential is the Euler one. 
Finally, in Section \ref{sec:conclusions} we discuss the similarities between the basis functions of Poisson's equation for the Euler problem and the corresponding ones
for the wave equation in the  Kerr metric.


\section{Euler's `two-centre' gravitational field}
\label{sec:euler}

\subsection{The analogue of Carter's constant in Newtonian gravity}

Particle motion in stationary and axisymmetric gravitational fields is characterized by a conserved energy $E$ and angular momentum $L_z$ along the symmetry axis. In this case the connection between the symmetries of the background field and the nature of the
conserved quantities is, more or less, obvious. The existence of an additional third integral is a more rare property, and is not related to any obvious symmetry of the gravitational field. In General Relativity, and more specifically in the case of the Kerr metric, the presence of such a conserved quantity--the Carter constant--is a consequence of the underlying presence of a rank-two Killing tensor. The Carter constant itself is known to be the sum of the total angular momentum $L^2$ plus
a spin-dependent term; its true meaning (to some extent) has been clarified only recently~\cite{rosquist}.

In Newtonian gravity there are very few potentials $V$ that are solutions of the vacuum Poisson equation $\nabla^2 V =0$ and 
which are endowed with a third integral. Within the general class of stationary and axisymmetric potentials the simplest one
possessing a third inegral is the superposition of a monopole-dipole field. Using standard spherical coordinates 
$\{r,\theta,\varphi\}$, this potential is (we set $G=c=1$ adopting the geometrized units of General Relativity)
\be
V_{\rm md} (\mathbf{r}) = -\frac{M}{|\mathbf{r}|} + \frac{\mathbf{d}\cdot\mathbf{r}}{|\mathbf{r}|^3} 
\label{dipole}
\ee
where $M$ is the mass of the source and $\mathbf{d}$ is its dipole moment. According to the Landau \& Lifshitz Mechanics textbook \cite{LLbook}, $V_{\rm md}$ is the most general potential in {\em spherical} coordinates that leads to a separable Hamilton-Jacobi equation for particle motion\footnote{Interestingly, on page 901 of Ref.~\cite{MTW} it is stated that the integrability of motion in the potential $V_{\rm md}$, inspired Misner to suggest Carter to look for a third integral in the Kerr metric.}. 

Another example of a potential with a third integral constant is provided by the famous solution found by Leonhard Euler in 1760 for
the problem of two fixed gravitating masses. This system consists of two static point masses $m_1$ and $m_2$ with a fixed distance 
$2a$ between them. Without loss of generality we can assume that the two masses are located on the $z$-axis 
(which thus becomes the symmetry axis) at symmetrical places around the origin.
The resulting potential is 
\be
V_{\rm E}=-\frac{m_1}{|\mathbf{r}-a \mathbf{\z}|}-\frac{m_2}{|\mathbf{r}+a \mathbf{\z}|} 
\quad \to \quad  
V_{\rm E} = -\frac{m_1}{r_1} -\frac{m_2}{r_2}
\label{euler}
\ee
where $\pm a \mathbf{\z}$ are the positional vectors of the two masses with respect to the origin 
(a `hat' denotes a unit vector) and $r_1, r_2$ are the distances of the given observation point from the
two masses. 

The third integral associated with point-particle motion in this potential is then
\be
Q = \frac{1}{2}\mathbf{h}_1 \cdot \mathbf{h}_2+
(m_2 \hat{\mathbf{r}}_2 - m_1 \hat{\mathbf{r}}_1) \cdot \mathbf{\hat{z}} a
\label{Qconst}
\ee
where $\hat{\mathbf{r}}_i$ is the unit vector from the $i$-th mass to the particle and $\mathbf{h}_i$ is the particle's reduced angular momentum with respect to the $i$-th mass; i.e. $\mathbf{h}_i=\mathbf{r}_i \times \mathbf{v}$ with $\mathbf{v}$ the
particle's velocity. It should be noted that the explicit form of $Q$ can be derived in more than one ways: for example, Lynden-Bell~\cite{lyndenbell} provides a straightforward derivation of this result, while Ref.~\cite{LLbook} arrives at the same result by separating the Hamilton-Jacobi equation.

In a recent paper~\cite{will10}, Will was able show that there is a {\em unique}\footnote{The uniqueness of this potential has been formally proven very recently, see Ref.~\cite{markakis}.} stationary/axisymmetric {\em and} reflection-symmetric Newtonian potential that possesses a third integral that is quadratic with respect to the momentum (as the Carter constant). 
It turns out that this special potential is identical to the reflection-symmetric case of the Euler solution (\ref{euler}), that is when $m_1 = m_2 = M/2$. For later convenience this potential will be denoted as $V_{\rm sE}$. Hence, 
\be
V_{\rm sE}=- \frac{M/2}{|\mathbf{r}-a \mathbf{\z}|} - \frac{M/2}{|\mathbf{r}+a \mathbf{\z}|}
\label{V2c}
\ee
Note that the additional requirement of reflection symmetry has eliminated the potential $V_{\rm md}$ (eqn.~(\ref{dipole})) 
from the list of potentials with a third integral since the dipole field lacks this property.

Will arrived at the potential $V_{\rm sE}$ by asking a question with bearing on the Kerr metric. 
Considering a general decomposition of a stationary, axisymmetric, and reflection-symmetric potential in multipole moments 
$M_l$ (where $l=0,1,2,\ldots$ is an integer),
\be
V = -\sum_{l = 0}^{\infty} \frac{M_{2l}}{|\mathbf{r}|^{2l+1}} P_{2l} (\cos\theta),
\label{Vexpan}
\ee
what is the relation (if any) obeyed by $M_{2l}$, in order for $V$ to lead to particle motion with a 
third integral/Carter-like constant? It turns out that a recurrence relation does exist between the moments and is given by
\be
M_l = \frac{M_2}{M} M_{l-2}\quad \to \quad  M_{2 l}=M \left( \frac{M_2}{M} \right)^l,
\label{moments}
\ee
where $M_0 = M$, $M_2$ is the quadrupole moment and so on (only the even-order moments $M_{2l}$ appear here, while the odd-order moments are all vanishing, $M_{2l+1}=0$, as a result of the imposed reflection symmetry).

Remarkably, this is the very same relation obeyed by the mass multipole moments of the Kerr metric, as defined by Hansen and Geroch~\cite{moments}. Exactly this relation, together with a similar one for the current-mass multipoles (which of course have no analogue in Newtonian gravity), completely characterizes the Kerr metric, and encapsulates the mathematical form of the famous `no-hair' theorem for Kerr black holes~\cite{MTW}.

The nature of the Newtonian source with the structure of eqn.~(\ref{moments}) is easily revealed if we  use in (\ref{Vexpan}) 
the following identity for the Legendre polynomials $P_l(\mu)$ (correcting a typo in \cite{will10}, in the generating Legendre function formula just above eqn.~(22) of that paper)
\bear
\sum_{l=0}^{\infty} t^{2 l} P_{2 l} (\mu) &=& 
\frac{1}{2} \left( \sum_{l=0}^{\infty} t^{l} P_{l} (\mu) +
\sum_{l=0}^{\infty} (-t)^{l} P_{l} (\mu) \right) \nonumber \\
&=&
\frac{1}{2}\left[(1-2 t \mu + t^2)^{-1/2}+(1+2 t \mu + t^2)^{-1/2}\right].
\eear
Then (\ref{Vexpan}) reduces to the potential $V_{\rm sE}$ of (\ref{V2c}) with\footnote{Based on intuition alone, one might have expected that the Newtonian potential closer to the structure of the Kerr metric would be the field of a circular ring distribution of matter, as a substitute of the ring singularity of Kerr black holes. However, the explicit calculation of the multipoles 
$M_{2l}= (-1)^l M b^{2l} (2l-1)!!/(2^l l!)$ (where $b$ is the ring's radius) reveals a clearly non-Kerr multipolar structure. 
This result is probably due to the fact that the usual spherical coordinates used to decompose a Newtonian potential into multipole moments are not exactly analogous to the Boyer-Lindquist coordinates in which the Kerr metric is usually described.}
\be
a^2 = \frac{M_2}{M}.
\ee
For a \emph{prolate} gravitational field, $M_2 > 0$, the distance $a$ is a real number and the resulting gravitational field source can be identified with Euler's two-centre system (two point masses $m_1=m_2=M/2$ located at $z=\pm a$). On the other hand, the \emph{oblate} field $M_2 <0$ (which is the one that corresponds to the multipolar structure of the Kerr metric) entails an imaginary distance\footnote{In order to avoid any misunderstanding we will keep the distance $a$ real, and add the imaginary unit $i$ whenever we discuss the oblate case.} $ia$. The mass moments then become
\bear
& M_{2 l}= M a^{2l}, \qquad &\mbox{(prolate)}
\\ \nonumber \\
& M_{2 l}= M (-a^2)^{l}, \qquad &\mbox{(oblate)}.
\eear

The connection between the Euler potential and the Kerr metric is clearly intriguing although there are still some distinct differences. The Kerr metric describes an oblate gravitational field, as one would naturally expect due to its intrinsic spin.
In contrast, the real space Euler potential is prolate by construction. This difference is reflected in the sign of the quadrupole moment $M_2$; Kerr's quadrupole moment being negative, while Euler's quadrupole moment is positive.

Opposite sign quadrupole moments would lead to a qualitatively different particle motion in the two fields. This difference 
can be annulled artificially by `rotating' the distance $a$ on the complex plane, i.e. $a \to ia$. 
Although this procedure cannot be produced by a suitable rearrangement of the masses in the physical space, the resulting potential is a legitimate, real-valued solution of the Poisson equation. This fact is obvious from the expansion~(\ref{Vexpan}), but it can also be derived by considering the symmetric Euler potential (\ref{V2c}) and changing $a$ to $ia$. Thus, after all, it is more appropriate to think of the oblate Euler potential (which has the same set of mass moments as Kerr) as the 'Kerr-surrogate' in 
Newtonian gravity (a more detailed discussion on the extraordinary similarities between the two fields will be presented elsewhere~\cite{AposHatzPapa}).

The oblate Euler solution can be interpreted as a two-centre system in the complex $\mathbf{a}$ space with the masses located at 
$\pm ia \mathbf{\z}$. We should note that the transformation $a \to i a $ should be accompanied by expressing vector amplitudes
(denoted as $|\ldots|$) according to the rule $| \mathbf{k} | \equiv \sqrt{\mathbf{k} \cdot \mathbf{k} }$.
This product is a complex function and in order to keep the square root single-valued square root we should place a branch cut along the negative real axis of the vector inner product.

With this complex-plane extension of the Euler solution, the general potential $V_{\rm E}$ assumes real values only in the symmetric case $m_1=m_2$ since then the two terms in (\ref{euler}) are complex conjugate to each other, $|\mathbf{r}-i  a \mathbf{\hat{z}}|= |(\mathbf{r}+i  a \mathbf{\hat{z}})|^*$.
After some lines of algebra the resulting oblate potential, henceforth denoted as $\tilde{V}_{\rm sE}$, takes the
real-valued form
\be
\tilde{V}_{\rm sE} = -\frac{M}{\sqrt{2} R^2} \sqrt{ R^2 + r^2-a^2 },
\label{VsEsym}
\ee
where $ R^2 = \sqrt{(r^2 -a^2)^2 +(2 a \mathbf{r}\cdot\mathbf{\hat{z}})^2} $.

A similar result holds for the Carter-like constant Q, eqn.~(\ref{Qconst}): assuming an oblate Euler field, 
the vectors $\hat{\mathbf{r}}_2$ and $\hat{\mathbf{r}}_1$ are complex conjugate with respect to each other;
this makes $\hat{\mathbf{r}}_2 - \hat{\mathbf{r}}_1$ a purely imaginary quantity. The same is true for the inner product 
$\mathbf{h}_1 \cdot \mathbf{h}_2$, since we have that $\mathbf{h}_1 = \mathbf{h}_2^* $. Given these properties, and despite the presence
of the imaginary parameter $ia$, the constant Q is a real number provided the system is symmetric, $m_1=m_2$.


\subsection{The separabillity of the Hamilton-Jacobi equation}
\label{sec:sphero}

Before moving to the main topic of this paper (the separability of the Newtonian wave equation) it is worth considering the separability of the Hamilton-Jacobi equation for point particle motion. The aim here is not to provide any new results but merely highlight some key points with relevance to our discussion.

As discussed in more detail in Ref.~\cite{LLbook}, the separability of the Hamilton-Jacobi equation in the gravitational field
of the Euler potential can be achieved in \emph{spheroidal} coordinates $\{\xi,\eta,\varphi\}$. The following discussion can be made
more compact if we simultaneously study both the prolate and oblate Euler fields. To this end we will have
to use on the same footing both prolate and oblate spheroidal coordinates: all formulae will be expressed in both coordinates; the upper (lower) sign will be reserved for prolate (oblate) coordinates, whenever such a discrimination is needed.

The spheroidal coordinates $\{\xi,\eta,\varphi\}$ are related to the standard spherical $\{r,\theta,\varphi \}$ and cylindrical $\{\varpi,z,\varphi \}$ coordinates as
\be
\varpi^2 = r_0^2 (\xi^2 \mp 1) (1-\eta^2), \qquad z = r_0 \xi \eta, \qquad  \varphi = \varphi
\label{sphero1}
\ee
and
 \be
r = r_0 \left[ \xi^2 \pm (\eta^2-1) \right]^{1/2}, 
\qquad \cos\theta = \frac{\xi\eta}{\left[ \xi^2 \pm (\eta^2-1) \right]^{1/2}}
\ee
where $r_0$ is a constant length-scale. The spheroidal coordinates span the domain  $ 1 \leq \xi < \infty$ (prolate coordinates),
$0 \leq \xi < \infty$ (oblate coordinates), $-1 \leq \eta \leq 1$ and $ 0 \leq \varphi \leq 2\pi$. In other words, $\xi$ can be viewed as a radial coordinate, $\eta$ as the cosine of a latitude, and  $\varphi$ as the usual azimuthal 
angle. 

The spheroidal system at hand is in a sense `adapted' to the two-centre problem. In terms of the radial distances $r_1,r_2$ from the two centres and identifying $r_0 = a$ we have
\bear
\xi = \frac{r_1+r_2}{2a} ~&,&~  \eta = \frac{r_2-r_1}{2a} \qquad \textrm{(prolate)} \nonumber \\
\xi = \frac{r_1+r_2}{2a}
~&,&~ 
\eta = \frac{r_2-r_1}{2 i  a}
\qquad \textrm{(oblate)}.
\label{sphero2}
\eear
Relating the spheroidal coordinates to the spherical coordinates used in the earlier discussion of the two-centre system
is a straightforward task in the prolate case. The oblate case, on the other hand, is somewhat more involved as a result of
using the imaginary distance $ia$. The coordinate relation is 
\be
\xi=\sqrt{\frac{R^2+(r^2-a^2)}{2 a^2}}, \qquad
\eta=\textrm{sgn}(\cos \theta) \sqrt{\frac{R^2-(r^2-a^2)}{2 a^2}}, 
\ee
where $\textrm{sgn}(x)$ is the usual sign function and $R$ was defined immediately below eqn.~(\ref{VsEsym}).
It is noteworthy the fact that the oblate system contains points where $\xi=0$. Given that $r_1 = r_2^*$, 
we can have $\mathbf{r} \cdot \mathbf{\hat{z}}=0$ and $r^2-a^2<0$. Hence on the whole equatorial disk region $r \leq a$, $r_1$ and $r_2$ are purely imaginary complex conjugates and $\xi=0$.  

According to Ref.~\cite{LLbook} the most general stationary and axisymmetric potential $V(\xi,\eta)$ that renders separable the Hamilton-Jacobi equation in spheroidal coordinates (referred as ``elliptic coordinates'' in the textbook) is 
\be
V_{\rm sep} = \frac{F(\xi) + G(\eta)}{\xi^2 \mp \eta^2}
\label{Vsep1}
\ee
where $F, G$ are arbitrary functions. Although Ref.~\cite{LLbook} obtains this result in prolate spheroidal coordinates, it is
actually true in oblate spheroidal coordinates as well. In the ensuing analysis of the Hamilton-Jacobi equation
the Carter-like constant $Q$ emerges as one of the separation constants. This result is consistent with the third integral 
derivation of Ref.~\cite{lyndenbell}.

If, in addition, the potential $V_{\rm sep}$ is required (as it should!) to solve the Poisson equation we then arrive at the
less general expression (see Appendix~\ref{app:Vsep} for the derivation)
\be
V_{\rm sep} = \frac{A \xi + B \eta}{\xi^2 \mp \eta^2}
\label{Vsep2}
\ee
where $A,B$ are constants. With the help of (\ref{sphero2}) we can see that, in fact, this potential is {\em identical} to the general Euler potential, given by eqn.~(\ref{euler}), with the identification
\bear
A = -\frac{m_1+ m_2}{a} ~&,&~  B = -\frac{m_1-m_2}{a} \qquad (\textrm{prolate}), \nonumber \\
A = -\frac{m_1+ m_2}{a} ~&,&~  B = - i  \frac{m_1-m_2}{a} \qquad (\textrm{oblate}). 
\eear
Once again we can see that only the symmetric system $m_1=m_2$ generates a real-valued oblate potential ($B=0$). This 
potential then automatically becomes reflection symmetric with respect to the $z=0$ plane.


\section{The wave equation in Newtonian gravity}
\label{sec:wave}

\subsection{Formulation and separability}

As we emphasized in the Introduction, the Kerr metric is special because (among other things) it allows the separability 
of both the Hamilton-Jacobi and Teukolsky partial differential equations, in the well-known Boyer-Lindquist 
coordinates~\cite{MTW}. Our objective of this Section is to show that the Euler gravitational field is equally special,
in the sense that it is the \emph{unique} Newtonian potential (with the symmetries of stationarity, axisymmetry, and reflection-symmetry) that leads to separable Hamilton-Jacobi and wave equations. 
 
Before showing this in detail, we first ought to clarify what we mean by `Newtonian wave equation'. 
We define this equation as\footnote{An alternative definition of the wave equation could be based on the 
(time-independent) Schr\"odinger equation -- it can be shown that also this equation is separable in the Euler field. 
However, we prefer working with our definition of the wave equation as it is second order with respect to the time-derivative and
it does not contain the Planck constant, which would otherwise be needed to balance the dimensions in the
corresponding Schr\"odinger equation. We thank the anonymous referee for pointing out this possibility.}
\be
\Box \Psi = -\kappa V \Psi  
\label{wave1}
\ee
where $\Psi(t,\mathbf{r})$ is a (massless) test-scalar field which we assume to be artificially coupled to the 
gravitational potential V and $\kappa$ is a suitable constant so that the equation is dimensionally balanced. 
Assuming hereafter a harmonic time-dependence $\Psi \sim e^{i\omega t}$ we have the corresponding time-independent equation
\be
\nabla^2 \Psi + \left( \frac{\omega^2}{c^2} - \kappa V \right) \Psi = 0.
\label{wave2}
\ee
For the moment we will use the usual SI units in order to clarify some yet undefined issues with respect to $\kappa$, 
but later on we will return to the geometrized units which we employed previously in the paper.
Since a Newtonian gravitational potential has the dimensions of velocity squared there are only two possible choices for
$\kappa$: (i) $\kappa=c^2/(G M)^2$ and (ii) $\kappa=\omega^2/c^4$, plus combination of those two like $(c/GM)(\omega/c^2)$. 
If we introduce an extra length scale in the 
problem, like later on when we will study the separablity of the wave equation in spheroidal coordinates, another possible
choice for $\kappa$ will show up; $\kappa=1/(a^2 c^2)$. Furthermore, combinations of the latter expression for $\kappa$
with the previous ones, like $\kappa=\omega/(a c^3)$, could also correctly balance the dimensions in eqn.~(\ref{wave2}). 
However, one has to bear in mind that all $a$-dependent choices are pathological in the limit $a \to 0$, because then 
the potential term is divergent and we end up with a singular wave equation in a Keplerian potential. A similar situation
arises for the choice $\kappa=c^2/(G M)^2$; the resulting wave equation is problematic in the limit of a freely propagating scalar field, i.e. $M \to 0$.

Hence, the only viable, singularity-free, choice  is $\kappa=\omega^2/c^4$ and is the one we will adopt
in the following analysis despite the fact that it alters our picture of a fixed background gravitational potential. 
At this point, after having clarified the various possibilities for $\kappa$, we can resume the use of
geometrized units $c=G=1$ (and set $\kappa=\omega^2$ when, later on, we compare the Newtonian wave equation in 
spheroidal coordinates with the corresponding equation for Kerr).

We begin our analysis of eqn.~(\ref{wave2}) by first attempting separation of the wave equation in spherical coordinates.
Using the multipolar expansion (\ref{Vexpan}) it is easy to see that only the monopole term $- M/r$ allows separabality. 
Hence from early on we are forced to consider the problem in alternative coordinate systems.

Motivated by the separability of the Hamilton-Jacobi equation in spheroidal coordinates, it makes sense to study the 
wave equation in the prolate/oblate coordinate system $\{\xi,\eta,\varphi\}$ (having set $r_0 = a$ as before). 
In these coordinates the Laplacian takes the form (recall that the upper (lower) sign corresponds to prolate (oblate) coordinates):
\be
\nabla^2 \Psi = \frac{1}{a^2(\xi^2 \mp \eta^2)} \Bigg [ \partial_\xi \{ (\xi^2 \mp1) \partial_\xi \Psi \}  
+ \partial_\eta \{ (1-\eta^2) \partial_\eta \Psi  \} \Bigg ] 
+ \frac{\partial^2_\varphi \Psi }{a^2 (\xi^2 \mp 1) (1-\eta^2)}.
\label{laplace}
\ee
We then look for a separable solution of the form
\be
\Psi = R(\xi) S(\eta) e^{im\varphi} .
\ee
Inserting this in (\ref{wave2}) we obtain,
\bear
&&\Bigg[ 
(\xi^2 \mp 1)  \frac{\partial_\xi^2 R}{R} + 2 \xi \frac{\partial_\xi R}{R} \mp \frac{m^2}{\xi^2 \mp 1} \Bigg]
+ \Bigg[
(1-\eta^2) \frac{\partial_\eta^2 S}{S} - 2 \eta \frac{\partial_\eta S}{S} - \frac{m^2}{1-\eta^2} \Bigg] \nonumber \\
&&+a^2 (\xi^2 \mp \eta^2) (\omega^2 -\kappa  V)=0 .
\eear
Clearly, as far as separability is concerned, the key term is $ (\xi^2 \mp \eta^2) V$.
The most general functional form of $V$ that accomplishes separability is obviously
\be
(\xi^2 \mp \eta^2) V = F(\xi) + G(\eta),
\label{sepa}
\ee 
where $F, G$ are arbitrary functions. This is nothing else but the potential $V_{\rm sep}$ 
of eqn.~(\ref{Vsep1}) that separates the Hamilton-Jacobi equation as well. We then end up with the following pair of ODEs:
\bear
&&\partial_\eta \left \{ (1-\eta^2) \partial_\eta S \right \} -\left [ \frac{m^2}{1-\eta^2}
\pm (\omega a \eta)^2  + \kappa a^2   G(\eta) +E \right ] S =0
\label{angular}
\\
\nonumber \\
&& (\xi^2  \mp1) \partial_\xi^2 R + 2 \xi \partial_\xi R +
 \left [ (a \omega \xi)^2 \mp \frac{m^2}{\xi^2 \mp 1} - \kappa a^2  F(\xi) 
-E \right ] R = 0  
\label{waves}
\eear
where $E$ is the angular separation constant.

Now, if we impose  the additional requirement $\nabla^2 V =0$, so that $V$ corresponds to a gravitational potential, 
we find (see Appendix~\ref{app:Vsep}) that separation of the wave equation is achieved only for the Euler gravitational potential
(and in the case of oblate coordinates only for the symmetrical one, if we insist on real-valued potentials):
\bear
V=V_{\rm E}  = -\frac{   (m_1+m_2) \xi + (m_2-m_1) \eta  }
{a(\xi^2 - \eta^2)}                           \qquad ({\rm prolate}) , \nonumber \\
V={\tilde V}_{\rm sE} = -\frac{   M \xi   }
{a(\xi^2 + \eta^2)}                          \qquad ({\rm oblate}) ,
\eear 
where $M=2 m_1=2 m_2$ in the oblate case.

In combination with the discussion of Section~\ref{sec:euler}, we have thus shown that the Euler field $V_{E}$ 
(and ${\tilde V}_{\rm sE}$), besides being the unique stationary and axisymmetric potential with a Carter-like constant and with 
a Kerr-like multipolar structure, it is also the only one that leads to fully separable Hamilton-Jacobi 
and scalar wave equations in  spheroidal coordinates. Based on these special properties we are compelled to conclude that the 
Euler field indeed qualifies as the \emph{Kerr analogue} of Newtonian gravity.

It is also interesting to determine the multipolar structure of $V_{\rm E}$ (or ${\tilde V}_{sE}$) in the spheroidal
coordinate system (we have already discussed the multipoles in spherical coordinates, see Eq.~(\ref{moments})).
To accomplish that we start by considering an axisymmetric potential of the form
\be
V(\xi,\eta) = \sum_n \cR_n (\xi) \cS_n (\eta)
\ee
Then from $\nabla^2 V=0$, written in spheroidal coordinates (either prolate or oblate), we obtain a pair of Legendre-type equations
\be
(1-\eta^2) \partial_\eta^2 \cS_n  - 2\eta \partial_\eta \cS_n + \lambda_n \cS_n = 0 ,
\ee
and
\be
(\xi^2 \mp 1) \partial_\xi^2 \cR_n + 2\xi \partial_\xi \cR_n -\lambda_n \cR_n = 0 , 
\ee
where $\lambda_n = n(n+1)$ is the corresponding angular eigenvalue. 
The solution that is reflection symmetric and asymptotically vanishing at infinity is
(after a trivial rescaling of the summation integer)
\bear
\label{Vexpan1}
V = \left\{ \begin{array}{ll}
\displaystyle\sum_{n = 0}^{\infty} \alpha_{2n} Q_{2n} (\xi) P_{2n} (\eta) ~&\textrm{(prolate potential)}  \\
&   \\
\displaystyle\sum_{n = 0}^{\infty} \alpha_{2n} i Q_{2n} (i  \xi) P_{2n} (\eta) ~&\textrm{(oblate potential)} 
\end{array} \right.
\eear
where $\alpha_k$ are constant coefficients and $P_k, Q_k$ are Legendre polynomials of the first and second kind, respectively \cite{ASbook}. The extra $i$ appearing in the oblate potential has been added to ensure a real-valued expression, 
assuming the coefficients $\alpha_{2n}$ are real (recall that $Q_{2n}$ is purely imaginary if its argument is purely imaginary).
Eqn.~(\ref{Vexpan1}) is the analogue of the multipolar expansion (\ref{Vexpan}) but in  spheroidal coordinates. 
The correspondence can be formalized if we use the large argument approximation for the Legendre polynomials either for 
real or imaginary argument~\cite{ASbook} 
\be
Q_n (\xi\to \infty) \approx \frac{n!}{(2n+1)!!} \xi^{-(n+1)} .
\ee
It then follows by a suitable dimensional rescalling of the terms in the series ($\xi$ is dimensionless) that 
\be
V(\xi \to \infty,\eta) \approx
\left\{ \begin{array}{ll}
-\sum_{n = 0}^{\infty} \cM_{2n} (a\xi)^{-(2n+1)} P_{2n}(\eta) ~&\textrm{(prolate potential)},   \\ 
& \\
-\sum_{n = 0}^{\infty} \cM_{2n} (a\xi)^{-(2n+1)} P_{2n}(\eta) ~&\textrm{(oblate potential)}, \\
\end{array} \right. 
\ee
where the spheroidal multipole moments are then defined as
\be
\cM_{2n} = -\frac{(2n)! \, a^{2n+1}}{(4n+1)!!} \alpha_{2n} .
\label{moments2}
\ee
As we show in Appendix~\ref{app:moments}, both prolate and oblate series share the same
coefficients $\alpha_{2n}$, namely,
\be
\alpha_{2n} = -(4n +1) \frac{M}{a} .
\ee
Hence, the spheroidal moments of the Euler potential are given in closed form by
\be
\cM_{2n} = \frac{ (2n)!}{(4n-1)!!} M a^{2n}, \qquad n = 0, 1, 2, \ldots 
\ee
with $\cM_0 = M, \cM_2 = 2Ma^2/3$, etcetera. Most noteworthy, these moments do \emph{not} obey the Kerr rule~(\ref{moments}) 
and they are all positive regardless of prolateness/oblateness. The latter property is not too surprising given that the 
two types of potential are build on a different set of base functions ($Q_{2l}(\xi)$ or $i Q_{2l}(i \xi)$, respectively), which means
that the same coefficient terms do not mean identical potentials. At the same time the deviation from the Kerr multipole moments
is due to the fact that the Newtonian multipolar expansion is coordinate-sensitive, in contrast to the coordinate-invariant formulation of the Geroch-Hansen relativistic moments.


\subsection{Wave equation: Euler vs Kerr}
\label{sec:KvsE}

Having established the separability of the wave equation we now can focus on a closer comparison of that equation against the Kerr
scalar wave equation. Based on our previous discussion, it makes sense to consider for this comparison 
the oblate reflection-symmetric Euler potential ${\tilde V}_{\rm sE}$ which is the one closest to Kerr as we argued above.

With the identification $V={\tilde V}_{\rm sE}$, the angular and radial equations (\ref{angular}) and (\ref{waves})
become
\bear
&&\partial_\eta \left \{ (1-\eta^2) \partial_\eta S \right \} -\left [ \frac{m^2}{1-\eta^2}
- (\omega a \eta)^2  + E \right ] S = 0,
\label{Seuler}
\\
\nonumber \\
&& (\xi^2+1) \partial_\xi^2 R + 2 \xi \partial_\xi R + \left [ (a \omega \xi)^2 + \frac{m^2}{\xi^2+1} 
+  \kappa M a \xi -E \right ] R = 0 .
\label{Reuler}
\eear
The first equation has exactly the same form as the angular equation of the spin $s=0$ (scalar) spheroidal harmonic
functions -- an equation produced during the separation of the scalar wave equation (i.e. the $s=0$ Teukolsky equation, see
eqns.~(4.9) and (4.10) of \cite{teuk73}) in Boyer-Lindquist coordinates $\{t,r_{\rm BL},\theta_{\rm BL},\varphi_{\rm BL}\}$. 
Matching the two equations requires that we identify $\eta=\cos\theta_{\rm BL}$, where $\theta_{\rm BL}$ is the Boyer-Lindquist
latitude coordinate.

The second equation resembles the corresponding $s=0$ radial Teukolsky equation without, however, being identical to it. 
We can nevertheless push the resemblance further by replacing  the dimensionless coordinate $\xi$ in (\ref{Reuler}) 
with the following function of the Boyer-Lindquist radial coordinate $r_{\rm BL}$:
\be
\xi=\tan \Phi(r_{\rm BL}),
\label{transf}
\ee
with 
\be
\Phi(r_{\rm BL})=a \int_{r_{\rm min}}^{r_{\rm BL}} \frac{dx}{\Delta (x)} , \qquad \Delta(x) = x^2 -2Mx + a^2 .
\label{Phi}
\ee
where $r_{\rm min}(M,a)$ is the radial $r_{\rm BL}$ value that corresponds to $\xi=0$. This is chosen as to make
$\Phi(\infty)=\pi/2$ so that both $\xi$ and $r_{\rm BL}$ become simultaneously infinite. With this substitution
$0 \leq \xi < \infty$, $r_{\rm min} \leq r_{\rm BL} < \infty$ while $\xi(r_{\rm BL})$ is a purely monotonic function.

Performing this coordinate transformation on eqn.~(\ref{Reuler}) we obtain
\be
\Delta (r_{\rm BL}) \frac{d^2 R}{dr^2_{\rm BL}}+ 2 (r_{\rm BL}-M) \frac{dR}{dr_{\rm BL}}+ F(r_{\rm BL}) R=0 ,
\label{ODE}
\ee
with
\be
F(r_{\rm BL})=\frac{a^2 ( \omega^2 a^2 \tan^2 \Phi + \kappa M a \tan\Phi + m^2 \cos^2 \Phi - E)}
{ \Delta (r_{\rm BL}) \cos^2 \Phi } .
\label{thefunction}
\ee
Eqn.~(\ref{ODE}) has the same `symbolic' structure as  the radial scalar Teukolsky equation 
in Boyer-Lindquist coordinates. The residual difference appears in the form of the 
$F(r_{\rm BL})$ function between the two cases. For the Kerr scalar wave equation this function is~\cite{teuk73}
\be
F_{\rm K}(r_{\rm BL})=\frac{ \ [(r_{\rm BL}^2+a^2_{\rm K})\omega -a_{\rm K} m \ ]^2 }{\Delta (r_{\rm BL})}
+2 a_{\rm K} m \omega - a^2_{\rm K} \omega^2 -E,
\ee
where $a_{\rm K}$ is the usual Kerr spin parameter. 

The functions $F$ and $F_{\rm K}$ are clearly different; this is not surprising given the fundamentally different
physical character of the Euler and Kerr fields: (i) Black holes have an event horizon which acts as an one-way
membrane for impinging waves. No such boundary exists in the Newtonian problem\footnote{It can be shown that $r_{\min}(M,a)$
is located outside the corresponding Kerr event horizon, $r_{\min}>M+\sqrt{M^2-a^2}$.}. 
(ii) The Kerr metric contains rotation-induced
frame-dragging and this has a distinct impact on prograde and retrograde propagating waves. This effect is encoded in the
linear $m$ terms in $F_{\rm K}$. The Euler field, being oblivious to the direction of orbital motion, 
cannot discriminate between prograde and retrograde motion, and that is why $F(r)$ features $m^2$ terms only.
The Kerr-Euler dissimilarity can also be exposed
by taking the double limit $a\to 0$, $a_{\rm K} \to 0$ at which the two fields collapse to the Schwarzschild and Kepler fields 
respectively, which are fundamentally different. 

All these differences between the Kerr and Euler fields, regardless of their nature, refer to strong-gravity properties. 
Hence, there is still hope that $F$ and $F_{\rm K}$ (and as a result the wave equations themselves) could approach each other asymptotically far from the gravitating source. In fact, this should be required since both fields should approach the
Keplerian potential at $r_{\rm BL} \to \infty$. Indeed, it is not too difficult to verify that in this limit we have 
$F,F_{\rm K} \to \omega^2 r^2_{\rm BL}$.

The next order asymptotic term of $F$ depends explicitly on the scale parameter $\kappa$. The choice $\kappa =4 \omega^2$ 
leads to the identical behaviour 
\be
F, F_{\rm K} \approx \omega^2 r^2_{\rm BL} + 2 M \omega^2 r_{\rm BL}  +{O}(r_{\rm BL}^0)
\ee
at large $r_{\rm BL}$.
However, this is the furthest one can push the match between $F$ and $F_{\rm K}$. Once higher order terms are introduced, the two 
functions begin to deviate.

To conclude this Section, we return to the two basic Euler wave equations (\ref{Seuler}) and (\ref{Reuler}) and briefly discuss
their solutions. As we mentioned earlier, the former equation is solved by the $s=0$ spin-weighted spheroidal harmonics~\cite{teuk73}. 
The latter equation is solved by the generalized radial spheroidal harmonics of the first and second kind
$\Xi_n^{m(1,2)}(\omega a, i  \kappa a M|i  \xi)$ (for a detailed discussion of these functions see~\cite{Falloon}). 
The computation of such special functions though, especially in the case of imaginary arguments,
is an intricate issue (cf. \cite{Liu}) and thus it will not be attempted here.


\section{Concluding remarks}
\label{sec:conclusions}

Analogies between diverse physical systems is a much cherished property of Nature. In this paper we have 
discussed one such analogy between an extremely important, from the astrophysical point of view, relativistic solution
of the vacuum Einstein equations and a simple Newtonian potential that was first studied more than two centuries ago. 
The Euler field is the unique Newtionian potential that shares the key properties of the Kerr metric, namely, the separability
of the Hamilton-Jacobi equation (which leads to a Carter-like constant) and the scalar wave equation. At the same time the Euler and
Kerr fields share the same recurrence relation for their respective mass multipole moments. 

As far as the wave equation in the Euler field is concerned we have shown that the equations produced after the separation of 
variables can be cast in a Kerr-like form. While the angular equations are in fact identical the same is not true
for the radial equations. The two equations coverge to each other only in the weak gravity asymptotic region of large radii. 
Closer to the gravitating source (strong gravity region) the deviation is significant, an expected result given the
absence of event horizons and frame-dragging in Newtonian gravity. 

Future work on the subject will attempt exploiting the striking Euler-Kerr analogy at a more practical level, for instance, 
by studying the orbital motion of particles in the Euler field, and using this analogy to infer orbital properties of realistic cases related to a Kerr black hole as a central object.


\section*{Acknowledgements}
This work was supported by the research funding program of I.K.Y. (IKYDA 2010).
KG is supported by the Ram\'{o}n y Cajal Programme of the Spanish Ministerio de Ciencia e Innovaci\'{o}n 
and by the German Science Foundation (DFG) via SFB/TR7. 

\appendix

\section{The solution of  $\nabla^2 V_{\rm sep} =0$}
\label{app:Vsep}

In this appendix we show that the only separable gravitational field in spheroidal (either prolate or oblate)
coordinates is the Euler two-center potential.
Considering the potential (\ref{Vsep1}), we assume the following expansions for the
free functions $F(\xi), G(\eta)$:
\be
F(\xi) = \sum_{k=-1}^{+\infty} q_k \xi^{-k}, \qquad  G(\eta) = \sum_{k=0}^{+\infty} p_k \eta^{k}
\ee
These power-series ensure that the two functions are well-behaved at spatial infinity $\xi \to \infty$ and on the
equatorial plane $\eta = 0$. Then 
\be
V_{\rm sep} = \frac{q_{-1}\xi}{\xi^2 \mp\eta^2} + \frac{1}{\xi^2 \mp\eta^2} \sum_{k=0}^{+\infty} 
\left ( q_k \xi^{-k} + p_k \eta^k \right ) .
\label{Vsep4}
\ee
The first term in this expression solves the Poisson equation identically (the Laplacian operator is
given in eqn.~(\ref{laplace})). The remaining terms lead to 
\begin{multline}
\nabla^2 V_{\rm sep} = 0 \quad \to \quad 
\sum_{k=0}^{+\infty} \Bigg [ (k-1) p_k \eta^{k-2} \left 
\{\, (k-2)\eta^4  
+\eta^2 [4-k \mp (2+k)\xi^2] \pm k \xi^2 \,  \right \}
\\
+ (k+1) q_k \xi^{-(k+2)} \left \{\,  (k+2)\xi^4 \mp \xi^2 
[k+4+(k-2)\eta^2 ] + k\eta^2  \, \right \}  \Bigg ] =0 .
\end{multline}
This series vanishes only if  $q_k =0$ for all $k \geq 0$ and $p_k=0$ for all $k \geq 0$ except of $k = 1$. As a result,
the full solution takes the desired Euler form
\be
V_{\rm sep} = \frac{q_{-1}\xi + p_1 \eta}{\xi^2 \mp\eta^2} 
\ee
where $q_{-1}, p_1$ are constants. 


\section{Multipole moments in spheroidal coordinates}
\label{app:moments}

In order to calculate the specific mass moments of $V_{\rm E}$ (or ${\tilde V}_{\rm sE}$) in spheroidal coordinates we 
first write 
\bear
V_{\rm E} &=& -\frac{M \xi}{a(\xi^2-\eta^2)} = \sum_{n = 0}^{\infty} \alpha_{2n} Q_{2n} (\xi) 
P_{2n} (\eta), \nonumber \\
{\tilde V}_{\rm sE} &=& -\frac{M \xi}{a(\xi^2 +\eta^2)} = \sum_{n = 0}^{\infty} \alpha_{2n} i Q_{2n} (i \xi) 
P_{2n} (\eta) .
\label{expan1}
\eear
The objective now is to calculate the corresponding coefficients $\alpha_{2n}$. To this end we take the product of these expressions 
with $P_{2l}(\eta)$ and integrate:
\be
-\frac{M}{a} \xi \int_{-1}^{+1} d\eta \frac{P_{2l}(\eta)}{\xi^2-\eta^2} =
 \frac{2 \alpha_{2l}}{4l+1} Q_{2l} (\xi), 
 \label{step1a}
\ee
\be
 -\frac{M}{a} \xi \int_{-1}^{+1} d\eta \frac{P_{2l}(\eta)}{\xi^2+\eta^2} =
 \frac{2 \alpha_{2l}}{4l+1} i Q_{2l} (i \xi) .
\label{step1b}
\ee
With the help of the integral representation of $Q_k$ (Eq.~8.3.3 of \cite{ASbook})
\be
Q_k (z) = \frac{1}{2} \int_{-1}^{+1} dt \frac{P_k(t)}{z-t}, \qquad  ({\rm for }~\mbox{Re}(z) > -1, \quad \mbox{Im}(z)=0),
\ee
and the reflection property $Q_k(z)=(-1)^{k+1} Q_k(-z)$,
the left-hand sides of (\ref{step1a}) and (\ref{step1b}) can be written as
\bear
-\frac{M}{2 a} \Bigg [ \int_{-1}^{+1} d\eta \frac{P_{2l}(\eta)}{\xi +\eta} +  
\int_{-1}^{+1} d\eta \frac{P_{2l}(\eta)}{\xi -\eta}  
\Bigg ] &=& -\frac{2 M}{a} Q_{2l}(\xi), \nonumber \\
\rm{and} && \nonumber \\
-\frac{M}{2 i a} \Bigg [ -\int_{-1}^{+1} d\eta \frac{P_{2l}(\eta)}{i \xi +\eta} -  
\int_{-1}^{+1} d\eta \frac{P_{2l}(\eta)}{i \xi -\eta}  \Bigg ] 
&=& \frac{2 M}{i a} Q_{2l}(i\xi) .
\label{step2}
\eear
Combining these with (\ref{step1a}) and (\ref{step1b}) leads to
\be
\alpha_{2l} = -(4l +1) M/a
\ee
which, remarkably, has the same form for both oblate and prolate potentials. 




\begin{thebibliography}{}

\bibitem{chandrabook}
S. Chandrasekhar, {\em The Mathematical Theory of Black Holes}, (Oxford Univ. Press, 1983)

\bibitem{teuk73}
S.A. Teukolsky, Astroph. J. {\bf 185}, 635 (1973)

\bibitem{galactic}
J.\ Binney and S.\ Tremaine, {\it Galactic dynamics} (Princeton University Press, Princeton, New Jersey, 1987)

\bibitem{Arnold}
V.~I.~Arnold, {\em Mathematical Methods of Classical Mechanics (Second Edition)} (Springer-Verlag, New York, 1989)

\bibitem{carter}
B. Carter, Commun. Math. Phys. {\bf 10}, 280 (1968)


\bibitem{stewart}
J.M. Stewart and M. Walker, Proc R. Soc. Lond. A {\bf 341}, 49 (1974)

\bibitem{lyndenbell}
D. Lynden-Bell, MNRAS {\bf 338}, 208 (2003)

\bibitem{will10}
C. Will, Phys. Rev. Lett. {\bf 102}, 061101 (2010)

\bibitem{keres67}
H. Keres, Soviet Phys. JETP {\bf 25}, 504 (1967)

\bibitem{israel70}
W. Israel, Phys. Rev. D {\bf 2}, 641 (1970)


\bibitem{LLbook}
L.D. Landau and E.M. Lifshitz, {\em Mechanics (Third Edition)}, (Butterworth-Heinemann, Oxford, 1976)


\bibitem{rosquist}
K. Rosquist, T. Bylund and L. Samuelsson, Int. Journal.of Mod. Phys. D {\bf 18}, 429 (2009)

\bibitem{markakis}
C. Markakis, e-print arXiv:1202.5228

\bibitem{MTW}
C.W. Misner, K.S. Thorne and J.A. Wheeler, {\it Gravitation}, (Freeman, New York, 1973)

\bibitem{moments}
R. Geroch, J. Math. Phys. {\bf 11}, 2580 (1970); \\
R. Hansen, J. Math. Phys., {\bf 15}, 46 (1974); \\
G. Fodor, C. Hoenselaers, and Z. Perj\'es, J. Math. Phys., {\bf 30}, 2252 (1989)

\bibitem{ASbook}
M. Abramowitz and I. A. Stegun, {\em Handbook of Mathematical Functions}, (Dover, New York, 1965)

\bibitem{AposHatzPapa}
T.A. Apostolatos, G. Pappas and K. Chatziioannou, paper in preparation 


\bibitem{Falloon}P. E. Falloon, {\em Theory and Computation of Spheroidal
Harmonics with General Arguments}, (Master Thesis, University of Western Australia, 2001)

\bibitem{Liu} J W Liu, {\em J. Math. Phys.}, {\bf 33}, 4026 (1992)



\end{thebibliography}
\end{document}